 \documentclass[aps, twocolumn, nopacs, superscriptaddress,amsmath]{revtex4}
\usepackage{dcolumn}
\usepackage{bm}
\usepackage{graphicx}
\usepackage{color}
\usepackage{subfigure}
\usepackage{hyperref}
\usepackage{latexsym}
\usepackage{amsthm}
\usepackage{amssymb}
\usepackage{braket}
\usepackage{bbm}
\usepackage{mathbbol}
\DeclareGraphicsExtensions{.jpg,.pdf, .mps, .png, .eps, .ps, .EPS,.gif}

\begin{document}
\def\be{\begin{equation}}
\def\ee{\end{equation}}

\def\bc{\begin{center}}
\def\ec{\end{center}}
\def\bea{\begin{eqnarray}}
\def\eea{\end{eqnarray}}
\newcommand{\avg}[1]{\langle{#1}\rangle}
\newcommand{\Avg}[1]{\left\langle{#1}\right\rangle}

\def\ie{\textit{i.e.}}
\def\etal{\textit{et al.}}
\def\m{\vec{m}}
\def\G{\mathcal{G}}

\newcommand{\gin}[1]{{\bf\color{blue}#1}}
\newcommand{\juan}[1]{{\bf\color{red}#1}}

\title{Higher-order  simplicial  synchronization of coupled topological signals }
\author{Reza Ghorbanchian}
\affiliation{School of Mathematical Sciences, Queen Mary University of London, London, E1 4NS, United Kingdom}
\author{Juan G. Restrepo\footnote{Correspoding author. Email:juanga@colorado.edu}}
\affiliation{Department  of  Applied  Mathematics,  University  of  Colorado  at  Boulder,  Boulder,  CO  80309,  USA}
\author{Joaqu\'{\i}n J. Torres\footnote{Correspoding author. Email:jtorres@onsager.ugr.es}}
\affiliation{Departamento de Electromagnetismo y Física de la Materia and Instituto Carlos I de Física Teórica y Computacional, Universidad de Granada, 18071, Granada, Spain}
\author{Ginestra Bianconi\footnote{Correspoding author. Email:ginestra.bianconi@gmail.com}}
\affiliation{School of Mathematical Sciences, Queen Mary University of London, London, E1 4NS, United Kingdom}
\affiliation{
The Alan Turing Institute, The British Library, London, United Kingdom}
\begin{abstract}
Simplicial complexes capture the underlying network topology and geometry of complex systems ranging from the brain to social networks. Here we show that algebraic topology is  a fundamental tool to capture the higher-order dynamics of simplicial complexes. In particular we consider topological signals, i.e., dynamical signals defined on simplices of different dimension, here taken to be nodes and links for simplicity. 
We show that coupling between signals defined on nodes and links leads to explosive topological synchronization in which phases defined on nodes synchronize simultaneously to phases defined on links at a discontinuous phase transition. We study the model on real connectomes and on simplicial complexes and network models. Finally, we provide a comprehensive theoretical approach that captures this transition on fully connected networks and on  random networks treated within the annealed approximation, establishing the conditions for observing a closed hysteresis loop in the large network limit.
\end{abstract}
\maketitle

\section{Introduction}
Higher-order networks \cite{giusti2016two,battiston2020networks,eliassirad,lambiotte} are attracting increasing attention as they are able to capture the many-body interactions of complex systems ranging from brain to social networks.
Simplicial complexes are higher-order networks that encode the network geometry and  topology of real datasets. Using simplicial complexes allows the network scientist to formulate new mathematical frameworks for mining  data \cite{otter2017roadmap,petri2013topological,massara2016network,sreejith2016forman,kartun2019beyond,katifori2} and for understanding these generalized network structures revealing the underlying deep physical mechanisms for emergent geometry \cite{wu,bianconi2017emergent,bianconi2016network,tadic1,tadic2} and for higher-order dynamics \cite{millan2020explosive,torres2020simplicial,reitz2020higher,
barbarossa2020topological,landry2020effect,skardal2019abrupt,skardal2019higher,
iacopini2019simplicial,taylor2015topological,lucas2020multiorder,zhang2020unified,skardal2020memory,lee,carletti2020dynamical,millan2018complex,mulas2020coupled,gambuzza2020master,millan2019synchronization}.
In particular, this very vibrant research activity is relevant in neuroscience to analyse real brain data and its profound relation to dynamics \cite{giusti2016two,severino2016role,sporns,giusti2015clique,reimann2017cliques,petri2013topological,tadic2} and in the study of biological transport networks \cite{katifori1,katifori2}.

\begin{figure*}[htb!]
\centering
\includegraphics[width=2.0\columnwidth]{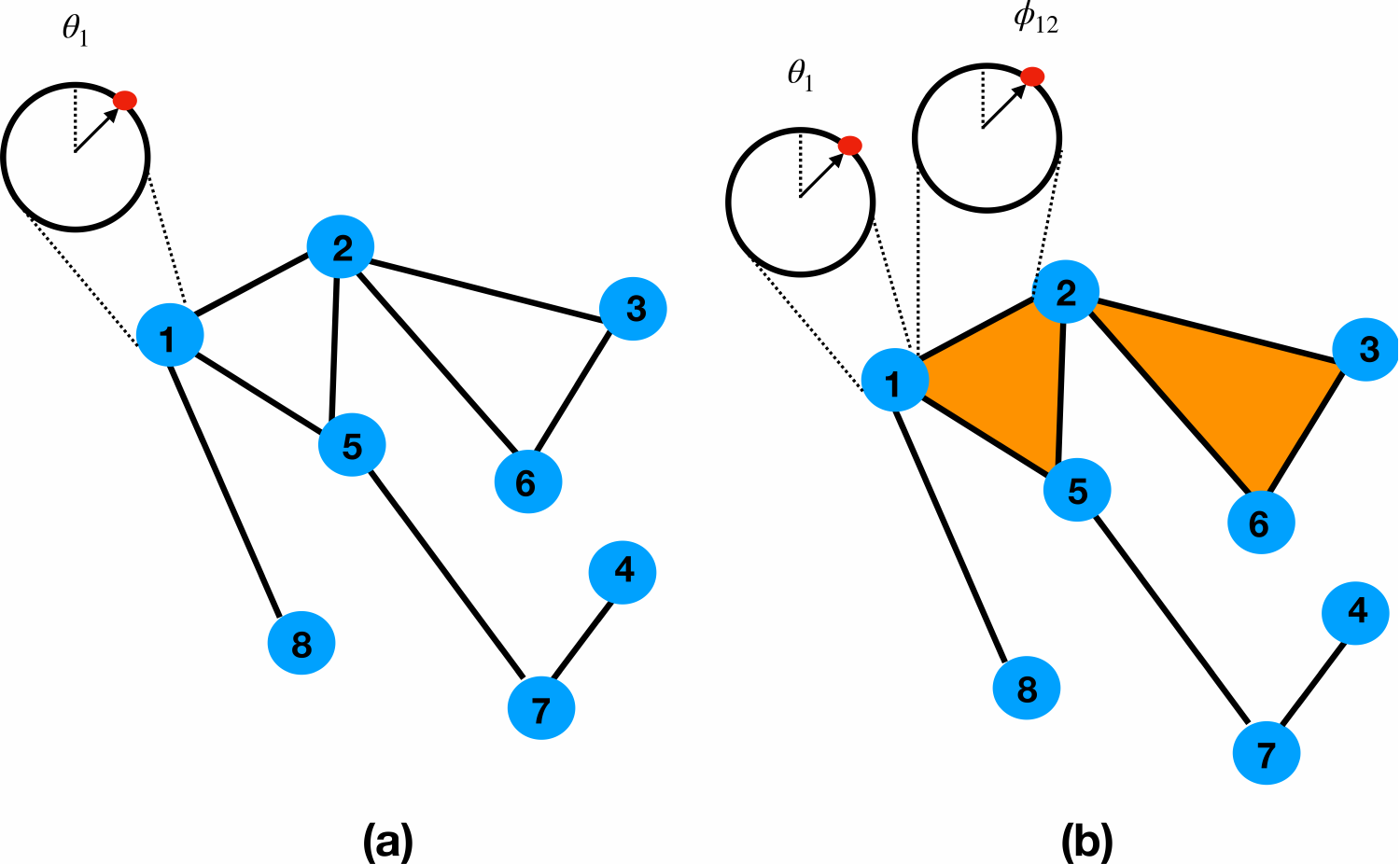}
\caption{{\bf Schematic representation of the Kuramoto and the topological Kuramoto model.} Panel (a) shows a network in which nodes sustain a dynamical variable (a phase) whose synchronization is captured by the Kuramoto model. Panel (b) shows a simplicial complex in which not only nodes but also links sustain   dynamical variables whose coupled synchronization dynamics is captured by the higher-order topological Kuramoto model.}
\label{fig:diagram}
\end{figure*}

In networks, dynamical processes are  typically defined over signals associated to the nodes of the network. In particular, the Kuramoto model \cite{strogatz2000kuramoto,boccaletti2018synchronization,rodrigues2016kuramoto,
restrepo2005onset,ott2008low}
 investigates the synchronization of phases associated to the nodes of the network. This scenario can change significantly in the case of simplicial complexes \cite{millan2020explosive,torres2020simplicial,barbarossa2020topological}.
In fact, simplicial complexes can sustain dynamical signals defined on simplices of different dimension, including nodes, links, triangles and so on, called {\em topological signals}. For instance, topological signals defined on links can represent fluxes of interest in neuroscience and in biological transportation networks.
The interest on topological signals is rapidly growing with new results related to signal processing  \cite{torres2020simplicial,barbarossa2020topological} and higher-order topological synchronization \cite{millan2020explosive,lee,note}.
In particular, higher-order topological synchronization \cite{millan2020explosive} demonstrates that topological signals (phases) associated to higher dimensional simplices can undergo a synchronization phase transition.   These results open a new uncharted territory for the investigation of higher-order synchronization.

 Higher-order topological signals defined on simplices of different dimension  can interact  with one another in non-trivial ways.
For instance in neuroscience the activity of the cell body of a neuron can interact with synaptic activity which can be directly affected by gliomes in the presence of brain tumors \cite{araque2014gliotransmitters}. 
In order to shed light on the possible phase transitions that can occur when topological signals defined on nodes and links interact, here we build on the mathematical framework of higher-order topological synchronization proposed in Ref.~\cite{millan2020explosive} and consider a synchronization model in which topological signals of different dimension are coupled. We focus in particular on the coupled synchronization of topological signals defined on nodes and links, but we note that the model can be easily extended to topological signals of higher dimension.
The reason why we focus on topological signals {defined on nodes and links} is three-fold. First of all we can have a better physical intuition of  topological signals defined on nodes (traditionally studied by the Kuramoto model) and links (like fluxes) that is relevant in brain dynamics \cite{araque2014gliotransmitters,deville_brain} and biological transport networks \cite{katifori1,katifori2}. Secondly,
although the coupled synchronization dynamics of nodes and links can be considered as a special case of coupled synchronization dynamics of higher-order topological signals on a generic simplicial complex,  this dynamics can be observed also on networks including only pairwise interactions.  Indeed nodes and links are the simplices that remain unchanged if  we reduce a  simplicial complex to its network skeleton. Since currently there is more availability of network data than simplicial complex data, this fact implies that the coupled dynamics studied in this work has wide applicability as it can be tested on any network data and network model. Thirdly, defining the coupled dynamics of topological signals defined on nodes and links can open new perspectives in exploiting the properties of the line graph of a given network which is the network whose nodes corresponds to the links or the original network \cite{evans}.

{In this work, we show} that by adopting a global adaptive coupling of dynamics inspired by Refs.~\cite{d2019explosive,zhang2015explosive,dai2020discontinuous}  the coupled synchronization dynamics of topological signals defined on nodes and links is explosive \cite{boccaletti2016explosive}, i.e., it occurs at a discontinuous phase transition in which the two topological signals of different dimension synchronize at the same time.
{We also illustrate} numerical evidence of this discontinuity on real connectomes and on  simplicial complex models including the configuration model of simplicial {complexes} \cite{courtney2016generalized}  and the non-equilibrium simplicial complex model called Network Geometry with Flavor \cite{bianconi2016network,bianconi2017emergent}.
We provide  a comprehensive theory of this phenomenon on fully connected networks offering a complete analytical understanding of the observed transition. This approach  can be extended to  random networks treated within the annealed network approximation. The analytical results reveal  that the investigated transition is discontinuous.
 
\section{Results}
\subsection{Higher-order topological Kuramoto model of topological signals of a given dimension }
Let us consider a simplicial complex $\mathcal{K}$ formed by $N_{[m]}$ simplices of dimension $m$, i.e., $N_{[0]}$ nodes, $N_{[1]}$ links, $N_{[2]}$ triangles, and so on.
In order to define the higher-order synchronization of topological signals  we will make use of algebraic topology (see the Appendix for a brief introduction) and specifically we  indicate with   ${\bf B}_{[m]}$ the $m$-th incidence matrix representing the $m$-th boundary operator. \\ The higher-order Kuramoto model generalizes the classic Kuramoto model to treat synchronization of topological signals of higher-dimension.
The classic Kuramoto model describes the synchonization transition for  phases 
\bea
\bm{\theta}=(\theta_1,\theta_2,\ldots \theta_{N_{[0]}})
\eea
associated to  nodes, i.e., simplices of dimension $n=0$ (see Figure $\ref{fig:diagram}$).
The Kuramoto model is typically defined on a network but it can treat also synchronization of the phases associated to the nodes of a simplicial complex.
Each node $i$ {has associated} an internal frequency $\omega_{i}$ drawn from a given distribution, for instance a normal distribution $\omega_i\sim \mathcal{N}(\Omega_0,1/\tau_0)$. In absence of any coupling, i.e., in absence of pairwise interactions, every node oscillates at its own frequency. However in a network or in a simplicial complex skeleton the phases associated to the nodes follow the dynamical evolution dictated by the equation    
\bea
 \dot{\bm \theta}=\bm \omega-\sigma  {\bf B}_{[1]}\sin \left({\bf B}_{[1]}^{\top}\bm\theta\right),
 \label{K0}
\eea
where here and in the following  we use the notation  $\sin({\bf x})$ to indicate the column vector where the sine function is taken element wise. Note that here we have chosen to write this system of equations in terms of the incidence matrix ${\bf B}_{[1]}$. However if we indicate with ${\bf a}$ the adjacency matrix of the network and with $a_{ij}$ its matrix elements, this system of equations is equivalent to 
\bea
\dot{\theta}_i=\omega_i+\sigma\sum_{j=1}^N a_{ij}\sin(\theta_j-\theta_i),
\eea
 valid for every node $i$ of the network.
 For coupling constant $\sigma=\sigma_c$ the Kuramoto model \cite{strogatz2000kuramoto,boccaletti2018synchronization,rodrigues2016kuramoto} displays a continuous phase transition  above which the order parameter 
 \bea
R_0&=&\frac{1}{N_{[0]}}\left|\sum_{i=1}^{N_{[0]}} e^{\mathbb{i}\theta_i}\right|
\label{op0}
\eea
is non-zero also in the limit $N_{[0]}\to \infty$.\\

 The higher-order topological Kuramoto model \cite{millan2020explosive} describes synchronization of phases associated to the $n$ dimensional simplices of a simplicial complex.
Although the definition of the model applies directly to any value of $n$, here we consider specifically the case in which the higher-order Kuramoto model is defined on topological signals (phases) associated to  the links
\bea
\bm{\phi}=(\phi_{\ell_1},\phi_{\ell_2},\ldots \phi_{\ell_{N_{[1]}}}),
\eea
where  $\phi_{\ell_r}$ indicates the phase
associated to the $r$-th link of the simplicial complex (see Figure $\ref{fig:diagram}$).
The higher order Kuramoto dynamics defined on simplices of dimension $n>0$ is the natural extension of the standard Kuramoto model defined by Eq.~(\ref{K0}). 
Let us indicate with $\tilde{\bm\omega}$ the {internal} frequencies associated to the links of the simplicial complex, sampled for example from a normal distribution,  $\tilde{\omega}_{\ell}\sim \mathcal{N}(\Omega_1,1/\tau_1)$. The higher-order topological Kuramoto model
is defined as
\bea
\dot{\bm{\phi}}&=&\tilde{\bm{\omega}}-\sigma  {\bf B}^{\top}_{[1]}\sin ({\bf B}_{[1]}\bm{\phi})-\sigma {\bf B}_{[2]}\sin ({\bf B}_{[2]}^{\top}\bm{\phi}).
\label{K1}
\eea
Once the synchronization dynamics is defined on higher-order topological signals of dimension $n$ (here taken to be $n=1$) an important question is whether  this dynamics can be projected on $(n+1)$ and $(n-1)$ simplices.
Interestingly, algebraic topology provides a clear solution to this question. Indeed for $n=1$, when the dynamics describes the evolution of phases associated {to the links}, one can consider the projection $\bm\phi^{[-]}$ and $\bm\phi^{[+]}$ respectively on nodes and on triangles defined as  
\bea
\bm\phi^{[-]}={\bf B}_{[1]}\bm\phi,\nonumber \\
\bm\phi^{[+]}={\bf B}_{[2]}^{\top} \bm\phi.
\eea 
Note that in this case ${\bf B}_{[1]}$ acts as a discrete divergence and ${\bf B}_{[2]}^{\top}$ acts as a discrete curl. Interestingly, since the incidence matrices satisfy ${\bf B}_{[1]}{\bf B}_{[{\text{\text{down}}}2]}={\bf 0}$ and ${\bf B}_{[2]}^{\top}{\bf B}_{[1]}^{\top}={\bf 0}$ (see Methods \ref{Ap0}) these two projected phases follow the uncoupled dynamics 
\bea
\dot{\bm \phi}^{[-]}&=&{\bf B}_{[1]}\tilde{\bm\omega}-\sigma {\bf L}_{[0]}\sin \bm\phi^{[-]},\nonumber \\
\dot{\bm \phi}^{[+]}&=&{\bf B}_{[2]}^{\top}\tilde{\bm\omega}-\sigma {\bf L}_{[2]}^{\text{\text{\text{down}}}}\sin \bm\phi^{[+]},\nonumber \\
\eea
where ${\bf L}_{[0]}={\bf B}_{[1]}{\bf B}_{[1]}^{\top}$ and  ${\bf L}_{[2]}^{\text{\text{down}}}={\bf B}_{[2]}^{\top}{\bf B}_{[2]}$.
These two projected dynamics undergo a continuous synchronization transition at $\sigma_c=0$ \cite{millan2020explosive} with order parameters 
\bea
R_1^{\text{down}}&=&\frac{1}{N_{[0]}}\left|\sum_{i=1}^{N_{[0]}} e^{\mathbb{i} \phi_i^{[-]}}\right|,\nonumber \\
R_1^{\text{up}}&=&\frac{1}{N_{[2]}}\left|\sum_{i=1}^{N_{[2]}} e^{\mathbb{i} \phi_i^{[+]}}\right|.
\label{op1}
\eea
In Ref. \cite{millan2020explosive} an adaptive coupling between these two dynamics is considered formulating the explosive higher-order topological Kuramoto model in which the topological signal follows the  set of coupled equations 
\bea
\dot{\bm{\phi}}&=&\tilde{\bm{\omega}}-\sigma R_1^{\text{up}} {\bf B}^{\top}_{[1]}\sin ({\bf B}_{[1]}\bm{\phi})\nonumber \\
&&-\sigma R_1^{\text{down}}{\bf B}_{[2]}\sin ({\bf B}_{[2]}^{\top}\bm{\phi}).
\eea
The projected dynamics on nodes and triangles are now coupled by the modulation of the coupling constant $\sigma$ with the order parameters $R_1^{\text{down}}$ and $R_1^{\text{up}}$, i.e. 
the two projected phases follow the coupled dynamics 
\bea
\dot{\bm \phi}^{[-]}&=&{\bf B}_{[1]}\tilde{\bm\omega}-\sigma R^{\text{up}}_1{\bf L}_{[0]}\sin \bm\phi^{[-]},\nonumber \\
\dot{\bm \phi}^{[+]}&=&{\bf B}_{[2]}^{\top}\tilde{\bm\omega}-\sigma R_1^{\text{down}} {\bf L}_{[2]}^{\text{down}}\sin \bm\phi^{[+]}.\nonumber \\
\eea
{This} explosive higher-order topological Kuramoto model has been shown in Ref.~\cite{millan2020explosive} to lead to a discontinuous synchronization transition on different models of simplicial complexes and on clique complexes of real connectomes.
 
\subsection{Higher-order topological Kuramoto model of coupled topological signals of different dimension}

Until now, we have captured synchronization occurring only among topological signals of the same dimension. However, signals of different dimension can be coupled to each other in non-trivial ways. 
In this work we will show how  topological signals of different dimensions can be coupled together leading to an explosive synchronization transition. Specifically we focus on the coupling of the traditional Kuramoto model [Eq.(\ref{K0})] to a higher-order topological Kuramoto model defined for phases associated to the links [Eq.(\ref{K1})].
The coupling between these two dynamics is here performed considering the modulation of the coupling constant $\sigma$ with the global order parameters of the node dynamics [defined in Eq. (\ref{op0})] and the link dynamics [defined in Eq. (\ref{op1})].
{Specifically, we consider two models denoted as Model NL (nodes and links) and model NLT (nodes, links, and triangles).} 
Model NL couples the dynamics of the phases of the nodes
 $\bm{\theta}$ and of the links $\bm{\phi}$ according to the following dynamical equations 
\bea
\dot{\bm{\theta}}&=&\bm{\omega}-\sigma R_1^{\text{down}}{\bf B}_{[1]}\sin ({\bf B}^{\top}_{[1]}\bm{\theta}),\label{thc}\\
\dot{\bm{\phi}}&=&\tilde{\bm{\omega}}-\sigma R_0 {\bf B}^{\top}_{[1]}\sin ({\bf B}_{[1]}\bm{\phi})-\sigma {\bf B}_{[2]}\sin ({\bf B}_{[2]}^{\top}\bm{\phi}).
\label{model1}
\eea
The projected dynamics for $\bm\phi^{[-]}$ and $\bm\phi^{[+]}$ {then obey }
\bea
\dot{\bm \phi}^{[-]}&=&{\bf B}_{[1]}\tilde{\bm\omega}-\sigma R_0{\bf L}_{[0]}\sin \bm\phi^{[-]},\label{phi-1} \\
\dot{\bm \phi}^{[+]}&=&{\bf B}_{[2]}^{\top}\tilde{\bm\omega}-\sigma  {\bf L}_{[2]}^{\text{down}}\sin \bm\phi^{[+]}.\label{phi+1}
\eea
Therefore the projection on the nodes ${\bm\phi}^{[-]}$ of the phases $\bm\phi$ associated to the links [Eq.~(\ref{phi-1})] is coupled to the dynamics of the phases $\bm{\theta}$ [Eq.~(\ref{thc})] associated directly to nodes. However the projection on the triangles ${\bm\phi}^{[+]}$ of the phases $\bm\phi$ associated to the links is independent of ${\bm\phi}^{[-]}$ and of $\bm\theta$ as well.
Model NLT also describes the coupled dynamics of topological signals defined on nodes and links but the adaptive coupling captured by the model is different. In this case the dynamical equations are taken to be  
\bea
\dot{\bm{\theta}}&=&\bm{\omega}-\sigma R_1^{\text{down}}{\bf B}_{[1]}\sin ({\bf B}^{\top}_{[1]}\bm{\theta}),\label{thc2}\\
\dot{\bm{\phi}}&=&\tilde{\bm{\omega}}-\sigma R_0 R_1^{\text{up}} {\bf B}^{\top}_{[1]}\sin ({\bf B}_{[1]}\bm{\phi})\nonumber \\
&-&\sigma R_1^{\text{down}}{\bf B}_{[2]}\sin ({\bf B}_{[2]}^{\top}\bm{\phi}).
\label{model2}
\eea
For Model NLT the projected dynamics for $\bm\phi^{[-]}$ and for $\bm\phi^{[+]}$ obey
\bea
\dot{\bm \phi}^{[-]}&=&{\bf B}_{[1]}\tilde{\bm\omega}-\sigma R_0R_1^{\text{up}}{\bf L}_{[0]}\sin \bm\phi^{[-]},\label{phi-2}  \\
\dot{\bm \phi}^{[+]}&=&{\bf B}_{[2]}^{\top}\tilde{\bm\omega}-\sigma  R_1^{\text{down}}{\bf L}_{[2]}^{\text{down}}\sin \bm\phi^{[+]}.\label{phi+2}  
\eea
Therefore, as in Model NL, the dynamics of the projection  ${\bm\phi}^{[-]}$ of the phases $\bm\phi$ associated to the links  [Eq.~(\ref{phi-2})] is coupled to the dynamics of the phases $\bm{\theta}$ associated directly to nodes [Eq.~(\ref{thc2})] and vice versa. Moreover, the dynamics of the projection  of the phases $\bm\phi$ on the triangles ${\bm\phi}^{[+]}$  [Eq.~(\ref{phi+2})]  is now also coupled with the dynamics of  ${\bm\phi}^{[-]}$ [Eq.~(\ref{phi-2})] and vice versa.
Here and in the following we will use the convenient notation (using the parameter $m$) to indicate both models NL and NLT with the same set of dynamical equations given by  
\bea
\dot{\bm{\theta}}&=&\bm{\omega}-\sigma R_1^{\text{down}}{\bf B}_{[1]}\sin ({\bf B}^{\top}_{[1]}\bm{\theta}),\label{thetai0}\\
\dot{\bm{\phi}}&=&\tilde{\bm{\omega}}-\sigma R_0 \left(R_1^{\text{up}}\right)^{m-1}{\bf B}^{\top}_{[1]}\sin ({\bf B}_{[1]}\bm{\phi})\nonumber \\
&&-\sigma \left(R_1^{\text{down}}\right)^{m-1} {\bf B}_{[2]}\sin ({\bf B}_{[2]}^{\top}\bm{\phi}),\label{phi0}
\eea
which reduce to Eqs.~(\ref{model1}) for $m=1$ and to Eqs.~(\ref{model2}) for $m=2$.

\begin{figure*}[htb!]
\centering
\includegraphics[width=2.0\columnwidth]{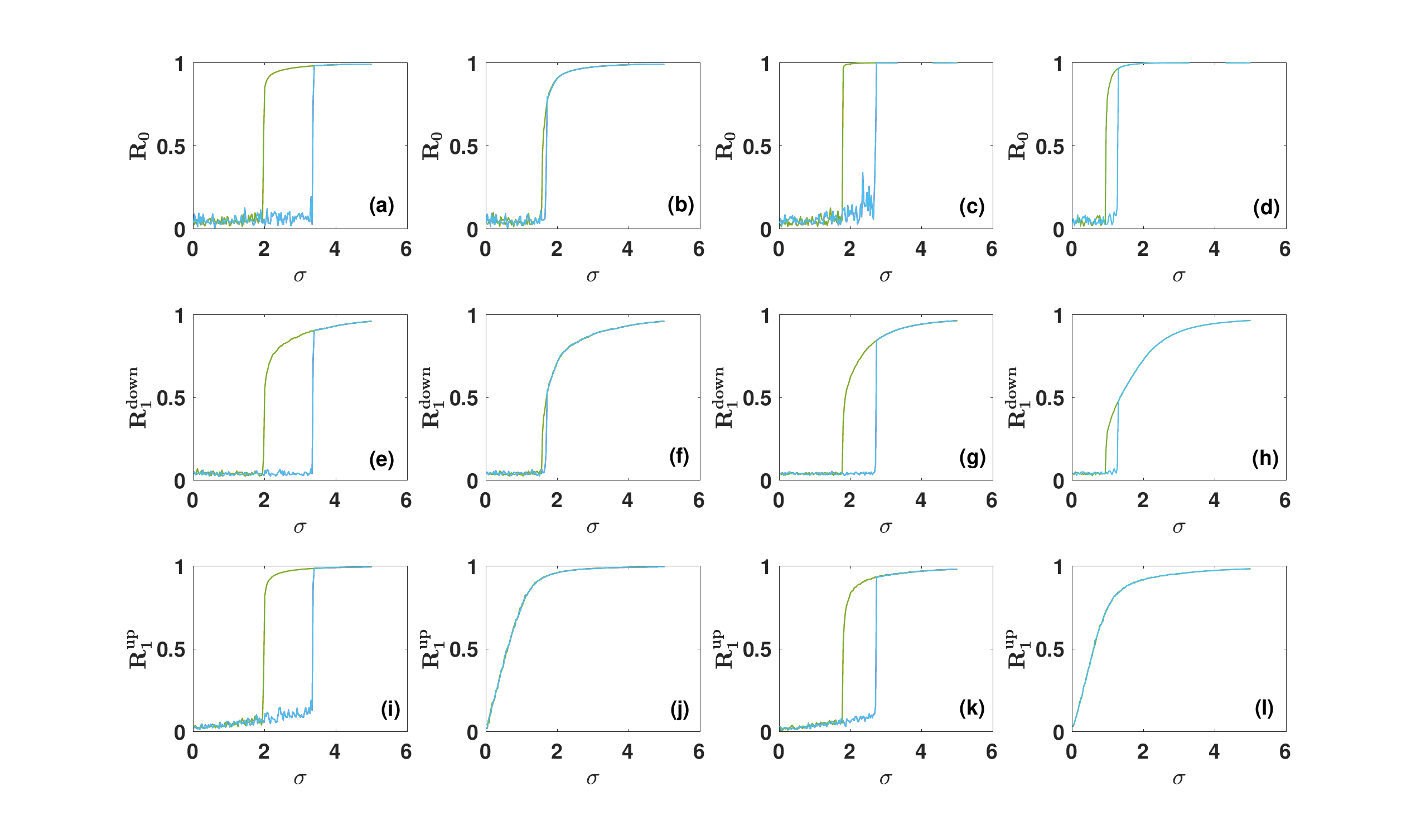}
\caption{{\bf The Higher-order topological synchronization models (Models NL and NLT) coupling nodes and links on simplicial complexes.} The order parameters $R_{0}$, $R_{1}^{\text{down}}$ and $R_1^{\text{up}}$ are plotted versus $\sigma$ for the higher-order topological synchronization Model NLT (panels (a)-(e)-(i) and (c)-(g)-(k)) and Model NL (panels (b)-(f)-(j) and (d)-(h)-(l)) defined over the Network Geometry with Flavor \cite{bianconi2016network} (panels (a)-(e)-(i) and (b)-(f)-(j)) and the configuration model of simplicial complexes \cite{courtney2016generalized} (panels (c)-(g)-(k) and (d)-(h)-(l)). The Network Geometry with Flavor on which we run the numerical results shown in (a) and (b) includes $N_{[0]}=500$ nodes and has flavor $s=-1$ and $d=3$. The configuration model of simplicial complexes on which we run the numerical results shown in (c) and (d) includes $N_{[0]}=500$ nodes and has generalized degree distribution which is power-law with exponent $\gamma=2.8.$ In both Model NL and in Model NLT we have set $\Omega_0=\Omega_1=2$ and $\tau_0=\tau_1=1$.}
\label{fig:model}
\end{figure*}

We make two relevant observations:
\begin{itemize}
\item
First, the proposed coupling between topological signals of different dimension can be easily extended to signals defined on higher-order simplices providing a very general scenario for coupled dynamical processes on {simplicial complexes.}
\item
Second, the considered coupled dynamics of topological signals defined on nodes and links can be also studied on networks with exclusively pairwise interactions where we assume that the number of simplices of dimension $n>1$ is zero.
Therefore in this specific case this topological dynamics can have important effects also on simple networks.
\end{itemize}
We have simulated {Model NL and Model NLT} on two main examples of simplicial complex models: the configuration model of simplicial complexes \cite{courtney2016generalized} and the Network Geometry with Flavor (NGF) \cite{bianconi2016network,bianconi2017emergent} (see Figure $\ref{fig:model}$). In the configuration model we have considered power-law distribution of the generalized degree with exponent $\gamma<3,$ and for the NGF model with have considered simplicial complexes of dimensions $d=3$ whose skeleton is a power-law network with exponent $\gamma=3$.
In both cases we observe an explosive synchronization of the topological signals associated to the nodes and to the links. On finite networks, the discontinuous transition  emerge together with the hysteresis loop formed by the forward and backward synchronization transition. However the two models display a notable difference. In Model NL we observe a discontinuity  for $R_0$ and $R_1^{\text{down}}$ at  a non-zero coupling constant $\sigma=\sigma_c$, however $R_1^{\text{up}}$ follows an independent transition at zero coupling (see Figure $\ref{fig:model}$, panels in the second and fourth column). In Model NLT, on the contrary, all order parameters $R_0$, $R_1^{\text{down}}$, and $R_1^{\text{up}}$ display a discontinuous transition occurring for the same non zero value of the coupling constant $\sigma=\sigma_c$ (see Figure $\ref{fig:model}$ panels in the first and third column).  This is a direct consequence of the fact that in Model NL the adaptive coupling leading to discontinuous phase transition only couples the phases $\bm\phi^{[-]}$ and $\bm\theta$, while for Model NLT the coupling involves also the phases $\bm\phi^{[+]}$.

\begin{figure*}[htb!]
\centering
\includegraphics[width=2.0\columnwidth]{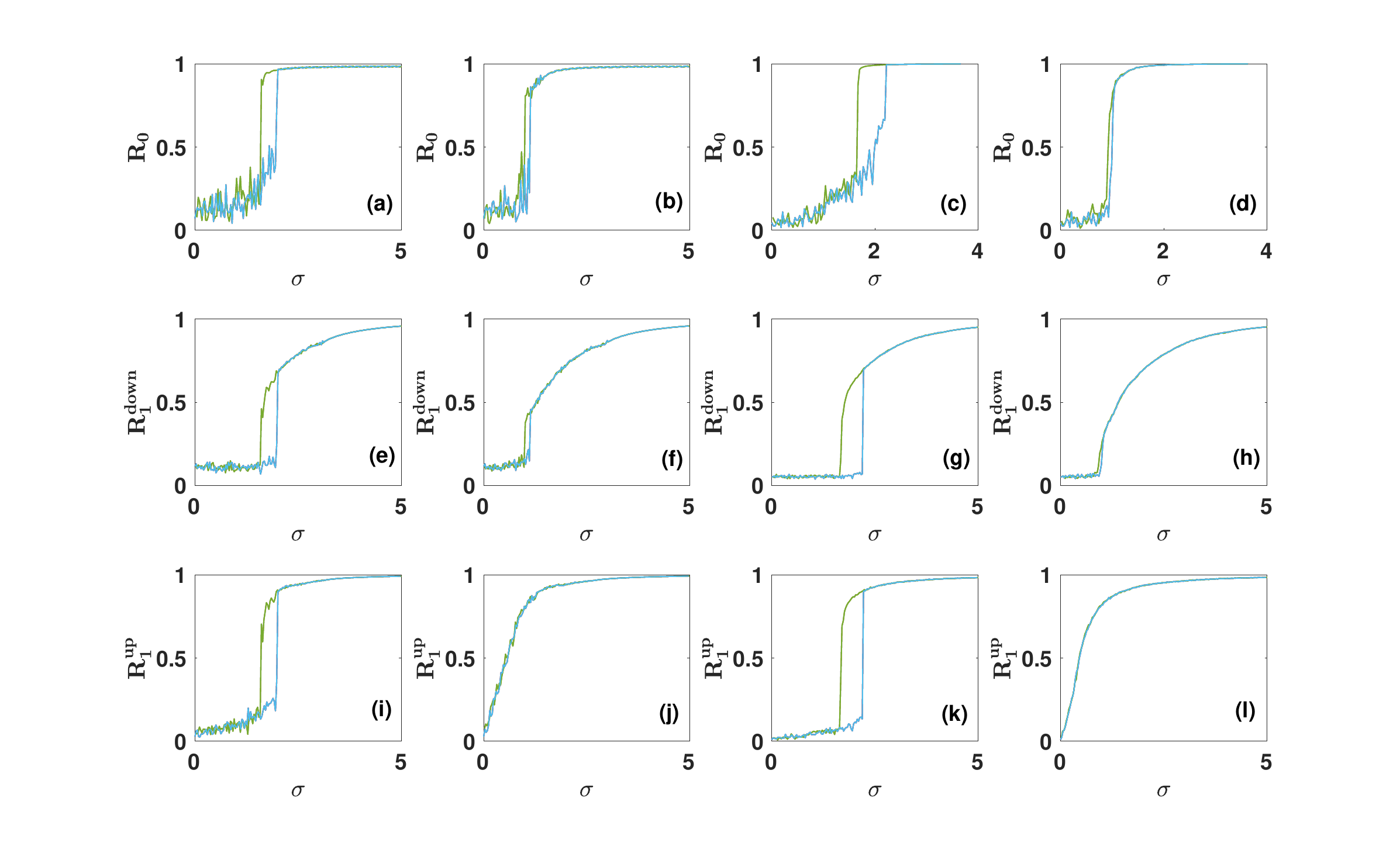}
\caption{{\bf The Higher-order topological synchronization models (Models NLT and NL) coupling nodes and links on real connectomes.} The order parameters $R_{0}$, $R_{1}^{\text{down}}$ and $R_1^{\text{up}}$ are plotted versus $\sigma$ on real connectomes. Panels (a)-(e)-(i) and (b)-(f)-(j) show the numerical results on the human connectome \cite{sapiens} for Model NLT and Model NL respectively. Panels (c)-(g)-(k) and (d)-(h)-(i) show the numerical results on the c. elegans connectome \cite{celegans} for Model NLT and Model NL respectively. In both Model NLT and in Model NL we have set $\Omega_0=\Omega_1=2$ and $\tau_0=\tau_1=1$. }
\label{fig:connectomes}
\end{figure*}

Additionally we studied both Model NL and Model NTL on two real  connectomes: the human connectome of Ref.~\cite{sapiens}  and the c. elegans connectome from Ref.~\cite{celegans} (see Figure $\ref{fig:connectomes}$). Interestingly also for these real datasets we observe that in Model NL the explosive synchronization involves only the phases $\bm\theta$ and $\bm\phi^{[-]}$ while in Model NLT we observe that also $\bm\phi^{[+]}$ undergoes an explosive synchronization transition at the same value of the coupling constant $\sigma=\sigma_c$.
 
\section{Discussion}

\subsection{Theoretical solution of the NL model}
As mentioned earlier the higher-order topological Kuramoto model coupling the topological signals of  nodes and  links can be defined on  simplicial complexes and on  networks as well.
In this section we exploit this property of the dynamics to provide an analytical understanding of the synchronization transition on uncorrelated random networks.

It is well known that the Kuramoto model is challenging to study analytically. Indeed the full analytical understanding of the model is restricted to the fully connected case, while on a generic sparse network topology the analytical approximation needs to rely on some approximations. A powerful approximation is the  annealed network approximation \cite{rodrigues2016kuramoto} which consists in approximating the adjacency matrix of the network with its expectation in a random uncorrelated network ensemble.
In order to unveil the fundamental theory that determines the coupled dynamics of topological signals described by the higher-order Kuramoto model here we  combine the annealed approximation with the  Ott-Antonsen method \cite{ott2008low}. 
This approach is able to capture the coupled dynamics of topological signals defined on nodes and links. In particular the solution found to describe the dynamics of topological signals defined on the links is highly non trivial and it is not reducible to the equations valid for the standard Kuramoto model. 
Conveniently, the calculations performed in the annealead approximation can be easily recasted in the exact calculation valid in the fully connected case previous a rescaling of some of the parameters. 
The analysis of the fully connected network reveals that the discontinuous sychronization transition of the considered model is characterized by a non-trivial backward transition with a well defined large network limit. On the contrary the forward transition  is highly dependent on the network size and  vanishes in the large network limit, indicating that the incoherent state remains stable for every value of the coupling constant $\sigma$ in the large network limit. This implies that on a fully connected network the NL model does not display a closed hysteresis loop as it occurs also for the model proposed in Ref. \cite{skardal2019abrupt}.
This scenario is here shown to extend also to sparse networks with finite second moment of the degree distribution while scale-free networks display a well defined hysteresis loop in the large network limit.

\subsection{Annealed dynamics}

For the dynamics  of the  phases $\bm\theta$ associated  to the nodes - Eq.~(\ref{thetai0}) - it is possible to proceed as in the traditional Kuramoto model \cite{ichinomiya2004frequency, lee2005synchronization,restrepo2005onset}. However  the annealed approximation for the  dynamics of the phases $\bm\phi$ defined in Eq. (\ref{phi0}) needs to be discussed in detail as it is not directly reducible to previous results.
To address this problem our aim is to directly define the annealed approximation for the dynamics of the projected variables $\bm\phi^{[-]}$ which, here and in the following are indicated as  
\bea
\bm \psi=\bm\phi^{[-]},
\label{projection0}
\eea
in order to simplify the notation. Moreover we will indicate with $N=N_{[0]}$ the number of nodes in the network or in the simplicial complex skeleton.
Here we focus on the NL Model defined on networks, i.e., we assume that there are no simplices of dimension two. 
We provide an analytical understanding of the  coupled dynamics of nodes and links in the NL Model by  determining the equations that capture the dynamics  in the annealed approximation and predict the value of the complex order parameters
\bea
{R}_0e^{\mathbb{i}\Theta}&=&\frac{1}{N}\sum_{i=1}^Ne^{\mathbb{i}\theta_i}, \nonumber \\
{R}_1^{\text{down}}e^{\mathbb{i}\Psi}&=&\frac{1}{N}\sum_{i=1}^Ne^{\mathbb{i}\psi_i},
\label{order}
\eea
(with $R_0,R_1^{\text{down}},\Theta$ and $\Psi$ real) as a function of the coupling constant $\sigma$.

We notice that  Eq.~(\ref{phi-1}), valid for Model NL, can be written as 
\bea
\dot {\bm{\psi}}&=&{\bf B}_{[1]}\tilde{\bm{\omega}}-\sigma R_0 {\bf L}_{[0]}\sin (\bm{\psi}).
\label{psi20}
\eea
This equation can be also written elementwise as
\bea
\dot {\psi}_i=\hat{\omega}_i+\sigma R_0 \sum_{j=1}^N a_{ij}\left[\sin(\psi_j)-\sin(\psi_i)\right],
\label{psi30}
\eea
where the vector $\hat{\bm \omega}$ is given by 
\bea
\hat{\bf\omega}={\bf B}_{[1]}\tilde{\bm{\omega}}.
\eea

Let us now consider in detail these frequencies in the case in which the generic internal frequency $\tilde{\omega}_{\ell}$ of a link  follows a Gaussian  distribution, specifically in the case in which ${\tilde\omega}_{\ell}\sim \mathcal{N}(\Omega_1,1/\tau_1)$  for every link $\ell$.
Using the definition of the boundary operator on a link it is easy to show that the expectation of $\hat{\omega}_i$ is given by  
\bea
\avg{\hat{\omega}_i}=\left[\sum_{j<i}a_{ij}-\sum_{j>i}a_{ij}\right]\Omega_1.
\label{avgomega}
\eea

Given that each node has degree $k_i$, the covariance matrix ${\bf C}$ is given by the graph Laplacian ${\bf L}_{[0]}$ of the network, i.e.
\bea
C_{ij}&=&\Avg{\hat{\omega}_i\hat{\omega}_j}_c=\sum_{\ell,\ell^{\prime}} \Avg{[{\bf B}_{[1]}\tilde{\bm\omega}]_i [{\bf B}_{[1]}\tilde{\bm\omega}]_j}_c\nonumber \\
&=&\frac{[{ L_{[0]}}]_{ij}}{\tau_1^2}=\frac{k_i\delta_{ij}-a_{ij}}{\tau_1^2},
\label{correlation}
\eea 
where we have indicated with $\Avg{\ldots}_c$ the connected correlation.
Therefore the
variance of $\hat{\omega}$ in the annealed approximation is 
\bea
\Avg{\hat{\omega}_i^2}_c=\avg{\hat{\omega}_i^2}-\avg{\hat{\omega}_i}^2=\frac{k_i}{\tau_1^2}.
\eea
Moreover, the projected frequencies are actually correlated and for $i\neq j$ we have 
\bea
\Avg{\hat{\omega}_i\hat{\omega_j}}_c=\Avg{\hat{\omega}_i\hat{\omega}_j}-\Avg{\hat{\omega}_i}\Avg{\hat{\omega}_j}=-\frac{a_{ij}}{\tau_1^2}.
\eea
It follows that the frequencies $\hat{\bm\omega}$ are correlated Gaussian variables with average given by Eq. (\ref{avgomega}) and correlation matrix given by the graph Laplacian.
The fact that the frequencies $\hat{\omega}_i$ are correlated is an important feature of the dynamics of $\bm\psi$ and, with few exceptions (e.g.,~\cite{skardal2015frequency}), this feature has remained relatively unexplored in the case of the standard Kuramoto model. 
Additionally let  us  note that the average of $\hat{\omega}$ over all the nodes of the network is zero. In fact
\bea
\sum_{i=1}^N \hat \omega_i = {\bf 1}^T \hat {\bm \omega} = {\bf 1}^T {\bf B}_{[1]}{\bm \omega} = 0,
\label{u}
\eea
where with ${\bf 1}$ we indicate the $N$-dimensional column vector of elements $1_i=1$.
By using the symmetry of the adjacency matrix, i.e. the fact that  $a_{ij} = a_{ji}$, Eq. (\ref{u}) implies that the sum of  $\dot \psi_i$ over all the nodes of the network is zero, i.e.
\bea
\sum_{i=1}^N \dot \psi_i &= \sum_{i=1}^N  \hat \omega_i + \sigma  R_0 \sum_{i,j} a_{ij}[\sin(\psi_j )- \sin(\psi_i)] = 0.\nonumber
\eea

We now consider the annealed approximation  consisting in substituting the adjacency matrix element $a_{ij}$ with its expectation in an uncorrelated network ensemble
\bea
a_{ij}\to \frac{k_ik_j}{\avg{k}N},
\label{annealed}
\eea
where $k_i$ indicates the degree of node $i$ and $\avg{k}$ is the average degree of the network. Note that the considered random networks can be both sparse \cite{anand2009entropy} or dense \cite{seyed2006scale} as long as they display the structural cutoff, i.e. $k_i\ll \sqrt{\avg{k}N}$ for every node $i$ of the network. 
In the annealed  approximation we can put 
\bea
\avg{\hat{\omega}_i}\simeq k_i\Omega_1 \left[1-2\sum_{j>i}\frac{k_j}{\avg{k}N}\right].
\label{homega_av0}
\eea
Also, in the annealed approximation the dynamical Eq.~(\ref{thetai0}) and Eq.~(\ref{psi20}) reduce to 
\bea
\dot{\bm \theta}&=&\bm\omega-\sigma R_1^{\text{down}}\hat{R}_0{\bf k}\cdot \sin(\bm{\theta}-\hat\Theta),\label{thetap} \\
\dot{\bm\psi}&=&\hat{\bm\omega}+\sigma  R_0 \hat{R}_1^{\text{down}}{\bf k}\sin\hat\Psi-\sigma R_0 {\bf k}\odot \sin \bm\psi, \label{psip}
\eea
where $\odot$ indicates the Hadamard product (element by element multiplication) and 
where   two auxiliary  complex order parameters are defined as 
\bea
\hat{R}_0e^{\mathbb{i}\hat\Theta}&=&\sum_{i=1}^N\frac{k_i}{\avg{k}N}e^{\mathbb{i}\theta_i}, \nonumber \\
\hat{R}_1^{\text{down}}e^{\mathbb{i}\hat\Psi}&=&\sum_{i=1}^N\frac{k_i}{\avg{k}N}e^{\mathbb{i}\psi_i},
\eea 
with $\hat{R}_0,\hat\Theta,\hat{R}_1^{\text{down}}$ and $\hat\Psi$ real.

\subsection{The dynamics on a fully connected network}

On a fully connected network in which each node has degree $k_i=N-1$ the dynamics of the NL Model is well defined provided its parameter are properly rescaled.
In particular we require a standard rescaling of the coupling constant with the network size, given by 
\bea
\sigma\to \sigma/(N-1)
\eea
which guarantees that the interaction term in the dynamical equations has a finite contribution to the velocity of the phases.
 
The Model NL on fully connected networks requires also some specific model dependent rescalings associated to the dynamics on networks. Indeed in order to have a finite expectation $\avg{\hat{\omega}_i}$ of the projected frequencies $\hat{\omega}_i$ and a finite of the covariance matrix $\bf C$, [given by Eqs.~(\ref{avgomega}) and (\ref{correlation}), respectively] we require that on a fully connected network both $\Omega_1$ and $\tau_1$ are rescaled according to  
\bea
\Omega_1 &\to& \Omega_1/N,\nonumber \\
\tau_1 &\to &\tau_1 \sqrt{N-1}.
\eea

Considering these opportune rescalings and noticing that the order parameters obey $\hat{R}_0=R_0$, $\hat{R}_1^{\text{down}}=R_1^{\text{down}}$,    $\Theta=\hat\Theta$, and $\Psi=\hat\Psi$, we obtain that Model NL  dictated by Eqs. (\ref{thetap})-(\ref{psip})  can be rewritten here as 
\bea
\dot{\bm \theta}&=&\bm\omega-\sigma R_1^{\text{down}}{R}_0 \sin(\bm{\theta}-\Theta),\label{thetapf} \\
\dot{\bm\psi}&=&\hat{\bm\omega}+\sigma  R_0 {R}_1^{\text{down}}\sin\Psi-\sigma R_0 \sin \bm\psi, \label{psipf}
\eea
with  $R_0,R_1^{\text{down}}, \Theta$ and $\Psi$ given by Eq. (\ref{order}) 
and 
\bea
C_{ij}=\Avg{\hat{\omega}_i\hat{\omega}_j}_c=\delta_{ij}-\frac{1}{N-1}.
\eea
\subsection{Solution of the dynamical  equations in the annealed approximation}

\subsubsection{General framework for obtaining the solution of the annealed dynamical equations}
In this section we will provide the analytic solutions for the order parameter of the higher-order topological synchronization studied within the annealed approximation, i.e., captured by Eqs.~(\ref{thetap}) and (\ref{psip}).
In particular first we will find an expression of  the order parameters $R_0$ of the dynamics associated to the nodes {(Eq.~(\ref{thetap}))} and subsequently in the next paragraph we will derive the expression for the order parameter $R_1^{\text{down}}$ associated to the projection on the nodes of the topological signal defined on the links (Eq.~{\ref{psip})). By combining the two results it is finally possible to uncover the discontinuous nature of the transition. 
\subsubsection{Dynamics of the phases of the nodes}
When we investigate Eq. (\ref{thetap}) we notice that this equation can be easily reduced to the equation for the standard Kuramoto model treated within the annealed approximation \cite{restrepo2005onset} if one performs a rescaling of  the coupling constant $\sigma $  $R_0\to \sigma$. Therefore we can treat this model similarly to the known treatment of the standard Kuramoto model \cite{restrepo2005onset,boccaletti2018synchronization,rodrigues2016kuramoto}.
Specifically, starting from Eq.~(\ref{thetap}) and  using a rescaling of  the phases $\bm{\theta}$ according to 
\bea
{\theta}_i\to {\theta}_i-\Omega_0 t, 
\eea
it is possible to show that we can set $\Theta=0$ and therefore Eq. (\ref{thetap}) reduces to the well-known annealed expression for the standard order Kuramoto model given by 
\bea
\dot{\bm \theta}&=&\bm\omega-\Omega_0 {\bf 1}-\sigma R_1^{\text{down}}\hat{R}_0{\bf k}\cdot \sin(\bm{\theta}).
\eea

Assuming that the system of {equations} reaches a steady state in which both $R_1^{\text{down}}$ and $\hat{R}_0$ become  time independent, the order parameters of this  system of equations in the coherent state $\hat{R}_0>0$ and $R_1^{\text{down}}>0$ can be found  to obey \cite{ichinomiya2004frequency,restrepo2005onset,boccaletti2018synchronization,boccaletti2016explosive} 
\bea
&&\hat{R}_0=\sum_{i=1}^N\frac{k_i}{\avg{k}N}\int_{|\hat{c}_i|<1} d\omega g(\omega) \sqrt{1-\left(\frac{\omega-\Omega_0}{\sigma k_i \hat{R}_0R_1^{\text{down}}}\right)^2},\nonumber \\
&&R_0=\frac{1}{N}\sum_{i=1}^N \int_{|\hat{c}_i|<1} d\omega g(\omega) \sqrt{1-\left(\frac{\omega-\Omega_0}{\sigma k_i \hat{R}_0R_1^{\text{down}}}\right)^2},\label{R0s_th}
\eea
where $\hat{c}_i$ indicates 
\bea
\hat{c}_i=\frac{\omega-\Omega_0}{\sigma k_j \hat{R}_0R_1^{\text{down}}}.
\eea
and $g(\omega)$ is the Gaussian distribution with expectation $\Omega_0$ and standard deviation $1$.

\subsubsection{Dynamics of the phases of the links projected on the nodes}
In this paragraph we will derive the expression of the order parameters $R_1^{\text{down}}$ and $\hat{R}_1^{\text{down}}$ which, together with Eqs.~(\ref{R0s_th}), will provide the annealed solution of our model.
To start with we assume that the frequencies $\hat{\bm\omega}$ are known. In this case we  can express the order parameters $R_1^{\text{down}}$ and $\hat{R}_1^{\text{down}}$ as a function of the 
probability  density function ${\rho}^{(i)}(\psi,t|\hat{\bm\omega})$ that node $i$ is associated to a projected phase of the link  equal to $\psi$. Since in the annealed approximation $\psi_i$ has a dynamical evolution dictated by Eq.~(\ref{psip}) the probability density function obeys the continuity equation
\bea
&&\partial_t {\rho}^{(i)}(\psi,t|{\hat{\bm\omega}})+\partial_{\psi}\left[{\rho}^{(i)}(\psi,t|{\hat{\bm\omega}}))v_{i}\right]=0\nonumber \\
\eea
with associated velocity $v_{i}$ given by 
\bea
v_{i}&=&\kappa_{i}-\sigma R_0 k_i\sin \psi_i,
\eea
where we have defined  $\kappa_i$ as 
\bea
\kappa_{i}&=&\hat{\omega}_i+\sigma k_i R_0\hat{R}_1^{\text{down}}\sin\hat\Psi.
\eea
In this case  the complex order parameters are given by 
\bea
\hat{R}_1^{\text{down}}e^{\mathbb{i}\Psi}&=&\sum_{i=1}^N\frac{k_i}{\avg{k}N}\int d\psi {\rho}^{(i)}(\psi,t|\hat{\bm\omega}) e^{\mathbb{i} \psi},\nonumber \\
{R}_1^{\text{down}}e^{\mathbb{i}\tilde{\Psi}}&=&\sum_{i=1}^N\frac{1}{N}\int d\psi {\rho}^{(i)}(\psi,t|\hat{\bm\omega}) e^{\mathbb{i} \psi}.
\label{R1_rho}
\eea

In order to solve the continuity equation we follow Ott-Antonsen \cite{ott2008low} and we express ${\rho}^{(i)}(\psi,t|\hat{\bm\omega})$ in the Fourier basis as 
\bea
{\rho}^{(i)}(\psi,t|\hat{\bm\omega})=\frac{1}{2\pi}\left\{1+\sum_{{m}=1}^{\infty}\hat{f}^{(i)}_{m}(\hat{{\omega}}_i,t) e^{\mathbb{i}{m} {{\psi}}}+c.c.\right\}.
\eea
Making the ansatz
\bea
\hat{f}_{m}^{(i)}(\hat{\omega}_i,t)&=&[b_i(\hat{\omega}_i,t)]^m 
\eea
we can derive the equation for the evolution of $b_i=b_i(\hat{\omega}_i,t)$ given by 
\bea
&&\partial_t b_i+\mathbb{i}b_i\kappa_{i}
+\sigma k_iR_0\frac{1}{2}(b^{2}_i-1)=0.
\label{dyn_0}
\eea
Since we showed before that the average value of $\dot \psi_i$ over nodes is zero, we look for non-rotating stationary solutions of Eq. (\ref{dyn_0}), $\partial_t b_i = 0$. As long as $R_0>0$ these stationary solutions are  given by
\bea
b_i&=&-\mathbb{i}d_i\pm \sqrt{1-d_i^2},
\eea
where $d_i$ is given by 
\bea
d_i=\frac{\hat{\omega}_i}{\sigma k_i R_0}+\hat{R}_1^{\text{down}}\sin\hat{\Psi}.
\eea
By inserting this expression into Eq.~(\ref{R1_rho}) we get  the expression of the order parameters given the projected frequencies $\hat{\bm\omega}$, in the coherent phase in which $R_0>0$
\bea
\hat{R}_1^{\text{down}}\cos\hat{\Psi}&=&
\sum_{i=1}^N\frac{k_i}{\avg{k}N}\sqrt{1-d_i^2}\theta(1-d_i^2),\nonumber \\
\hat{R}_1^{\text{down}}\sin\hat{\Psi}&=&\sum_{i=1}^N\frac{k_i}{\avg{k}N}\left\{\sqrt{d_i^2-1}\chi(d_i)+d_i\right\},\nonumber \\
{R}_1^{\text{down}}\cos\Psi&=&
\sum_{i=1}^N\frac{1}{N}\sqrt{1-d_i^2}\theta(1-d_i^2),\nonumber \\
{R}_1^{\text{down}}\sin\Psi&=&\sum_{i=1}^N\frac{1}{N}\left\{\sqrt{d_i^2-1}\chi(d_i)+d_i\right\},
\label{comp0}
\eea
where, indicating by $\theta(x)$  the Heaviside function, we have defined \bea
\chi(d_i)=[-\theta(d_i-1)+\theta(-1-d_i)].
\eea  
Finally, if the projected frequencies $\hat{\bm{\omega}}$ are not known we can average the result over the marginal frequency distribution of the projected frequency $\hat{\omega}_i$ given by $G_i(\hat{\bm\omega})$ getting 
\begin{widetext}
\bea
\hat{R}_1^{\text{down}}\cos\hat{\Psi}&=&
\sum_{i=1}^N\frac{k_i}{\avg{k}N}\int_{|d_i|\leq 1}d\hat{ \omega}_iG_i(\hat{ \omega}_i)\sqrt{1-\left(\frac{\hat{\omega}_i}{\sigma R_0k_i}+\hat{R}_1^{\text{down}}\sin \hat{\Psi}\right)^2}\nonumber ,\\
\hat{R}_1^{\text{down}}\sin\hat{\Psi}&=&-\sum_{j=0}^N\frac{k_i}{\avg{k}N}\int_{d_i>1}d\hat{\omega}_i G_i(\hat{ \omega}_i)\sqrt{\left(\frac{\hat{\omega}_i}{\sigma R_0k_i}+\hat{R}_1^{\text{down}}\sin \hat{\Psi}\right)^2-1}\nonumber \\
&&+\sum_{i=1}^N\frac{k_i}{\avg{k}N}\int_{d_i<-1}d\hat{ \omega}_iG_i(\hat{ \omega}_i)\sqrt{\left(\frac{\hat{\omega}_i}{\sigma R_0k_i}+\hat{R}_1^{\text{down}}\sin \Psi\right)^2-1}\nonumber \\
&&+\sum_{i=1}^N\frac{k_i}{\avg{k}N}\int_{-\infty}^{\infty}d\hat{ \omega}_iG_i(\hat{ \omega}_i) \left(\frac{\hat{\omega}_i}{\sigma R_0k_i}+\hat{R}_1^{\text{down}}\sin \Psi\right),\nonumber\\
R_1^{\text{down}}\cos \Psi&=&
\sum_{i=1}^N\frac{1}{N}\int_{|d_i|\leq 1}d\hat{ \omega}_iG_i(\hat{ \omega}_i)\sqrt{1-\left(\frac{\hat{\omega}_i}{\sigma R_0k_i}+\hat{R}_1^{\text{down}}\sin \hat{\Psi}\right)^2}\label{comp},
\eea
\end{widetext}
and an analogous equations for $R_1^{\text{down}} \sin(\Psi)$ (not shown). 
We note that in the case of distributions $g(\omega)$ and $G_i(\hat \omega)$ that are symmetric around their means  the above equations always admit the solution $\Psi = \hat \Psi = 0$.  Such values of the phases are also confirmed by direct numerical integration of the NL model.
\begin{figure*}[htb]
\centering
\includegraphics[width=1.8\columnwidth]{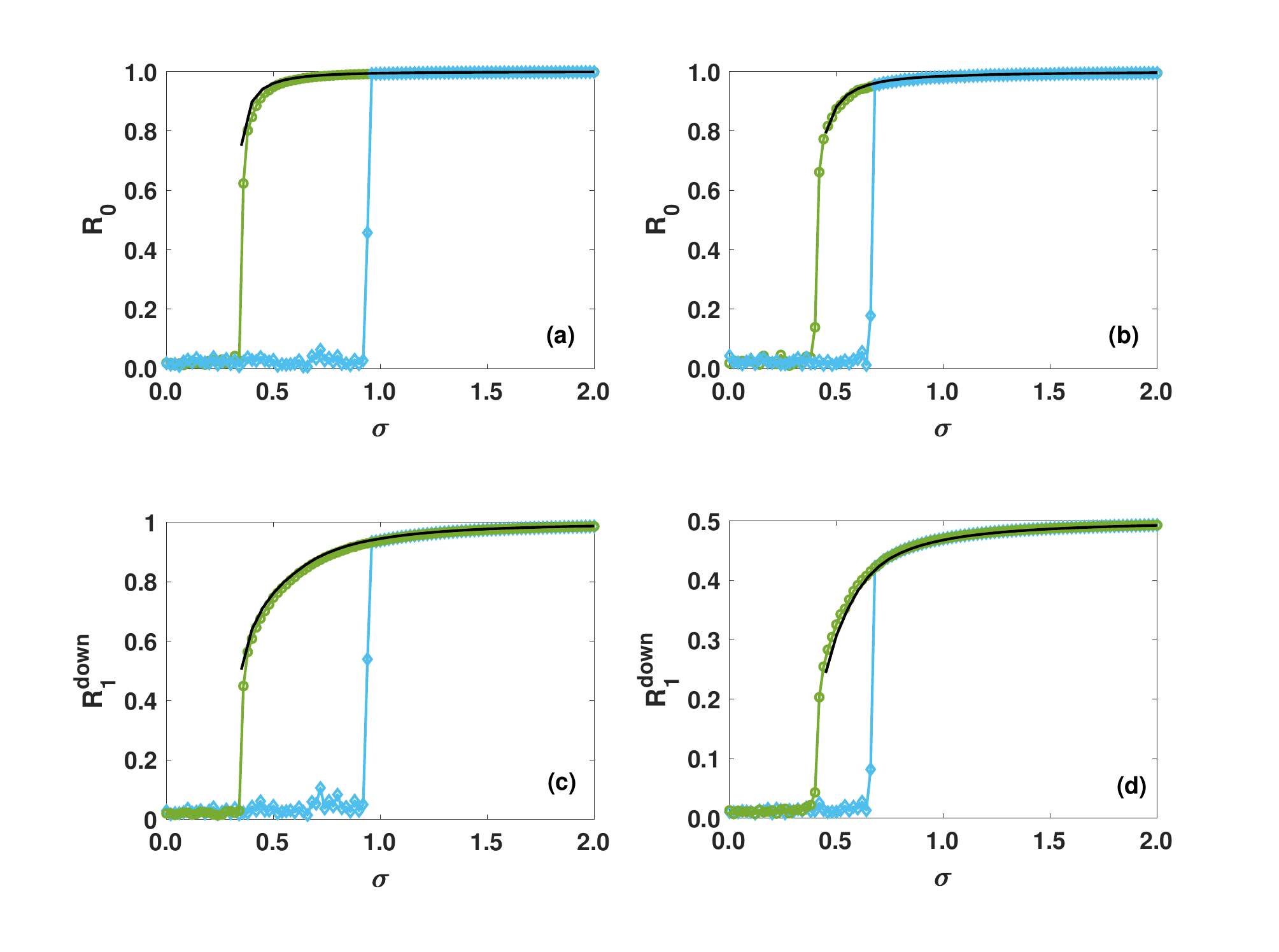}
\caption{{\bf Comparison between the simulation results of the NL Model and its solution in the annealed approximation}
The order parameters $R_0$ and $R_1^{\text{down}}$ of the NL Model are shown as a function of $\sigma$ for a Poisson network with average degree $c=12$ and for an uncorrelated  scale-free network with minimum degree $m=6$ and power-law exponent $\gamma=2.5$. Both networks have $N=1600$ nodes. The symbols indicate the simulation results for  the forward (cyan diamonds) and the backward (green circles) synchronization transition. The solid lines indicate the analytical solution for the backward transition obtained by integrating Eq.~(\ref{comp0}).}
\label{fig:annealed}
\end{figure*}
These equations together with Eqs.~(\ref{R0s_th}) capture the steady-state behavior of the higher-order Kuramoto model coupling topological signals defined on nodes and links within the annealed approximation in the coherent synchronized phase. 
Note that by derivation, these equations cannot capture the asynchronous phase  which is instead always a trivial solution of the dynamical equations corresponding to  $R_0=R_1^{\text{down}}=0$.
Finally we observe that for the NL Model as well as  for the standard Kuramoto model on random networks, it is expected  that the annealed approximation is more accurate for networks that are connected and are sufficiently dense.

To illustrate the applicability of the theoretical analysis, we consider two examples of connected networks with $N=1600$ nodes: a Poisson network with average degree $c=12$ and an uncorrelated  scale-free network with minimum degree $m=6$ and power-law exponent $\gamma=2.5$ In Fig.~\ref{fig:annealed} we compare the values of $R_0$, $R_1^{\text{down}}$ obtained from direct numerical integration of Eqs.~(\ref{thetai0}) and (\ref{psi30})  and the steady state solutions obtained from the numerical solution of Eqs.~(\ref{comp0}). 
The backward transition is fully captured by our theory, while the next paragraphs will clarify the theoretical expectations for the forward transition.  

\subsection{Solution on the fully connected network}

The integration of Eq.~(\ref{comp}) requires the knowledge of the marginal distributions $G_i(\hat{\omega})$ which does not have in general a simple analytical expression.
However, in the fully connected networks with Gaussian distribution of the internal frequency of nodes and links this calculation simplifies significantly. Indeed, when the   link frequencies are sampled from a  Gaussian distribution with mean $\Omega_1/N$ and  standard deviation  $1/(\tau_1\sqrt{N-1})$,  the marginal frequency distribution $G_i(\hat{\omega})$ of the internal frequency $\hat{\omega}_i$ of a node $i$ in a fully connected network is given by  (see Methods  for details)
\bea
G_i(\hat{\omega})&=& \frac{\tau_1}{\sqrt{2\pi/\bar{c}}}\exp\left[-\tau_1^2\bar{c}\frac{(\hat{\omega}_i-\Avg{\hat{\omega}_i})^2}{2}\right],
\label{marginal}
\eea
where $\bar c=\frac{N}{N-1}$. By considering $\Omega_0=\Omega_1=\Avg{\hat{\omega}_i}=0,$ and performing a direct integration of Eqs.~(\ref{comp})   we obtain (see Methods section for details) the closed system of equations for $R_0$ and $R_1^{\text{down}}$ 
\bea
&1 = \sigma   R_1^{\text{down}} h\left(\sigma^2  R_0^2 (R_1^{\text{down}})^2\right),\nonumber \\
&R_1^{\text{down}} = \sigma R_0 \tau_1 \sqrt{\bar{c}} h\left(\sigma^2 \tau_1^2 R_0^2 \right),\label{sys}
\eea
where the scaling function $h(x)$ is given by 
\bea
h(x) = \sqrt{\frac{\pi }{2}} e^{-x/4} \left[I_0\left(\frac{x}{4}\right)+I_1\left(\frac{x}{4}\right)\right],
\eea
with $I_0$ and $I_1$ indicating the modified Bessel functions.
The numerical solution of Eqs.~(\ref{sys})  reveals the following  picture: for low values of $\sigma$, only the incoherent solution $R_0  = R_1^{\text{down}} = 0$ exists. At a positive value of $\sigma$, two solutions of Eqs.~(\ref{sys}) appear at a bifurcation point, with the upper solution corresponding to a stable synchronized state and the lower solution to an unstable synchronized solution. For larger values of $\sigma$, the values of $R_0$ and $R_1^{\text{down}}$ corresponding to the upper solution approach one (full phase synchronization), while those for the lower solution approach zero asymptotically, thus indicating that the incoherent state never loses stability. 
Indeed, it can be easily checked (see Methods for details) that for large $\sigma$ the unstable solution of Eqs. (\ref{sys})  has  asymptotic behavior   
\bea
R_0&=&\sigma^{-2}J_0,\nonumber \\
R_1^{\text{down}}&=&{\sigma}^{-1}J_1,
\label{AsyFC}
\eea
with $J_0$ and $J_1$ constants given by
\bea
J_0&=&\left[\frac{\pi}{2}\right]^{-2}\left[G(0)g(0)\right]^{-1},\\
J_1&=&\left[g(0)\frac{\pi}{2}\right]^{-1}.
\label{JFC}
\eea
Therefore the unstable branch approaches the trivial solution $R_0=R_1^{\text{down}}=0$ only asymptotically for $\sigma\to \infty$. This implies that the trivial solution remains stable for every possible value of $\sigma$ although as $\sigma$ increases it describes the stationary state of an increasingly smaller set of initial conditions.

\begin{figure*}[htb]
\centering
\includegraphics[width=2.0\columnwidth]{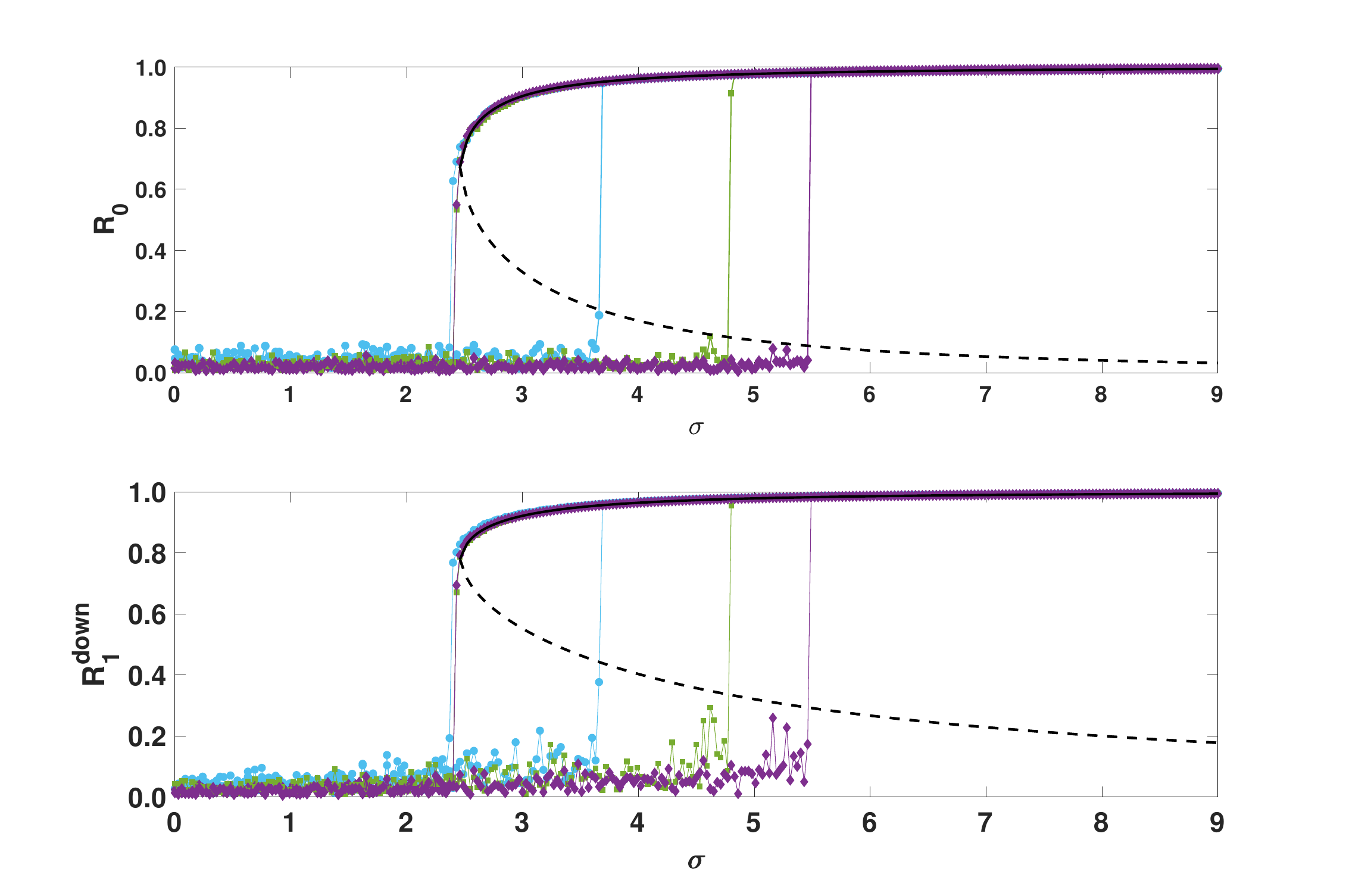}
\caption{{\bf The backward and the forward discontinuous phase transition  on fully connected networks}  The order parameters $R_0$ (circles) and $R_1^{\text{down}}$ (squares) are plotted as a function of the coupling constant $\sigma$  on a fully connected network. The solid and the dashed lines indicate the stable branch and the unstable branch predicted by  Eqs.(\ref{sys}). Simulations (shown as data point) are here obtained by integrating numerically  Eqs. (\ref{thetap}) and (\ref{psip}) for a fully connected network of $N=500$ (cyan circles), $N=1000$ (green squares), and $N=2000$ (purple diamonds)  with $\Omega_0=\Omega_1=0$ and (rescaled) $\tau_0=\tau_1=1$. The backward transition is perfectly captured by the theoretical prediction and is affected by finite size effects very marginally.  The forward transition is instead driven by stochastic fluctuations and moves to higher values of $\sigma$ as the network size increases.}
\label{fig:fully_connected}
\end{figure*}

 This scenario is confirmed by  numerical simulations (see Figure \ref{fig:fully_connected}) showing that  
the backward transition is captured very well by our theory and does not display notable finite size effects.
The forward transition, instead, displays remarkable finite size effects. Indeed, as $\sigma$ increases, the system remains in the incoherent state until it explosively synchronizes at a positive value of $\sigma$ and reaches the stable synchronized branch.
However the incoherent state is stable in the limit $N \to \infty$, and this forward transition is the result of finite size fluctuations that  push the system above the unstable branch, causing the observed explosive transition. This is consistent with the fact that for larger values of $N$, which have smaller finite size fluctuations, the system remains in the incoherent state for larger values of $\sigma$.

Therefore, while a closed hysteresis loop is not present in the NL model defined on fully connected networks, we observe fluctuation-driven hysteresis, in which finite-size fluctuations of the zero solution drive the system towards the synchronized solution, creating an effective hysteresis loop.

\subsection{Hysteresis  on homogeneous and scale-free networks}

In this section we discuss how the scenario found for the fully connected network can be extended to random networks with given degree distribution.
We will start from the self-consistent Eqs.~(\ref{comp}) obtained within the annealed approximation model. These equations display a saddle point bifurcation with the emergence of two non-trivial solutions describing a stable and an unstable branch of these self-consistent equations. These solutions always exist in combination with the trivial solution $R_0=R_1^{\text{down}}=0$ describing the asynchronous state. 
Two scenarios are possible:  either the unstable branch converges to the trivial solution only in the limit $\sigma\to\infty$ or it converges to the trivial solution at a finite value of $\sigma$.
In the first case, the scenario is the same as the one observed for the fully connected network, and the trivial solution remains stable for any finite value of $\sigma $. In this case the forward transition is not obtained in the limit $N\to\infty$ and the transition observed on finite networks is only caused  by finite size effects. In the second case the trivial solution loses its stability at a finite value of $\sigma$. Therefore the forward transition is not subjected to strong finite size effects and we expect to see a forward transition also in the $N\to\infty$ limit.
in order to determine which network topologies can sustain a non-trivial hysteresis loop we expand Eqs.~(\ref{comp}) for $0<R_0\ll1$, $0<\hat{R}_0\ll 1$, and $0<R_1^{\text{down}}\ll1$ under the hypothesis that the distributions $g(\omega)$ and $G_i(\hat{\omega}) $  are symmetric and unimodal.
Under these hypothesis it is easy to show that  Eqs.~(\ref{comp}) predict an unstable solution in which $R_0$ and $R_1^{\text{down}}$ scale with $\sigma$ according to
\bea
R_0&=&\sigma^{-2}J_0,\nonumber \\
R_1^{\text{down}}&=&{\sigma}^{-1}J_1,
\eea
with $J_0$ and $J_1$ constants given by
\bea
J_0&=&\Avg{k}\left[\frac{\pi}{2}\frac{\Avg{k^2}}{\Avg{k}}\right]^{-2}\left[ g(\Omega_0)\frac{1}{N}\sum_i k_iG_i(\Avg{\hat{\omega}_i})\right]^{-1},\nonumber\\
J_1&=&\left[g(\Omega_0)\frac{\pi}{2}\frac{\Avg{k^2}}{\Avg{k}}\right]^{-1}.
\eea
As long as the network does not have  vanishing  $J_0$ and $J_1$  the unstable branch converges to the trivial solution $R_0=R_1^{\text{down}}$ only in the limit $\sigma\to\infty$.
This happens for instance for Gaussian distribution of the internal frequency of the links and converging second moment $\avg{k^2}$ of the degree distribution. 
However, when the second moment diverges, i.e., the network is scale free with $\avg{k^2}\to \infty$ as $N\to\infty$, then $R_0$ and $R_1$ can converge to the trivial solution $R_0=R_1^{\text{down}}=0$ also for finite $\sigma$.
This analysis  suggests that the scenario described for the fully connected network remains valid for sparse (connected) networks as long as the degree distribution does not have a diverging second moment, while a stable hysteresis loop can be observed for scale-free networks.

\section{Conclusions}
Until recently  the synchronization phenomenon has been explored only in the context of topological  signals associated to the nodes of a network. However, the growing interest in simplicial complexes opens the perspective of investigating synchronization of higher order topological signals, for instance associated to the links of the discrete networked structure. Here we uncover how topological signals associated to nodes and links can be coupled to one another giving rise to an explosive synchronization phenomenon involving both signals  at the same time.
The model has been tested on real connectomes and on  major examples of simplicial complexes (the configuration model \cite{courtney2016generalized} of simplicial complex and the Network Geometry with Flavor \cite{bianconi2016network}).
Moreover, we provide an analytical  solution of this model  that provides a theoretical understanding of the mechanism driving the emergence of this discontinuous phase transition and the mechanism responsible for the emergence of a closed hysteresis loop. 
This work can be extended in different directions including the study of the   de-synchronization dynamics of this coupled higher-order synchronization and the duality of this model with the same model defined on the line graph of the same network.

\section*{Acknowledgements}
This work is partially founded by SUPERSTRIPES Onlus. This research utilized Queen Mary’s Apocrita HPC facility, supported by QMUL Research-IT.
http://doi.org/10.5281/zenodo.438045. G.B. acknowledge support from the Royal Society  IEC\textbackslash NSFC \textbackslash 191147.  J.J.T. acknowledges financial support from the Spanish Ministry of Science and Technology, and the Agencia Espa\~nola de Investigaci\'on (AEI) under grant FIS2017-84256-P (FEDER funds) and from the Consejer\'ia de Conocimiento, Investigaci\'on y  Universidad, Junta de Andaluc\'ia and European Regional Development Fund, Refs.~A-FQM-175-UGR18 and SOMM17/6105/UGR.

\section*{Author contributions}
All authors have contributed in the design of the project, in the numerical implementations of the algorithm, the theoretical derivations and the writing of the manuscript.

\section*{Code Availability}
All codes are available upon request to the Authors.
\section*{Data Availability}
The connectome network dataset used in this study are freely available: the Homo sapiens dataset comes from Ref. \cite{sapiens} the C.elegans dataset comes from Ref. \cite{celegans}.

\section*{Competing interests}
The authors declare no competing interests.

\bibliographystyle{apsrev4-1}
\bibliography{references}

\begin{thebibliography}{61}
\bibitem [{not()}]{note}%
  \BibitemOpen
  \href@noop {} {}\bibinfo {note} {Note that here higher-order refers to the
  higeher-order interactions existing between topological signals and not to
  higher-order harmonics.}\BibitemShut {Stop}%
\end{thebibliography}
\begin{center}
{\Large \bf METHODS}
\end{center}

\section{Simplicial complexes and higher order Laplacians}
\label{Ap0}
\subsection{Definition of simplicial complexes}
Simplicial complexes represent higher-order networks whose interactions include two or more nodes. These many-body interactions are captured by {\em simplices}.
 A $n$-dimensional {\it simplex} $\alpha$ is a set of $n+1$ nodes
\bea
\alpha=[i_0,i_1,\dots,i_{n}].
\eea 
For instance a node is a $0$-dimensional simplex, a link is a $1$-dimensional simplex, a triangle is a $2$-dimensional simplex, a tetrahedron is a $3$-dimensional simplex, and so on. A {\em face} of a simplex is the simplex formed by a proper subset of the nodes of the original simplex. For instance the faces of a tetrahedron are $4$ nodes, $6$ links and $4$ triangles.  A {\em simplicial complex} is a set of simplices closed under the inclusion of the faces of each simplex. Any simplicial complex can be {reduced} to its {\em simplicial complex skeleton}, which is the network formed  by the simplicial complex nodes and links.
Simplices have a relevant topological and geometrical interpretation and constitute the topological structures studied by discrete algebraic topology.
Therefore representing the many-body interactions of a complex system with a simplicial complex opens the very fertile opportunity to use  the tools of algebraic topology \cite{otter2017roadmap,ghrist2014elementary} to study the topology of  the system under investigation. In this {work} we show that algebraic topology can also shed significant light on the role that topology has on higher-order synchronization.
\begin{figure*}[htb!]
\centering
\includegraphics[width=2.0\columnwidth]{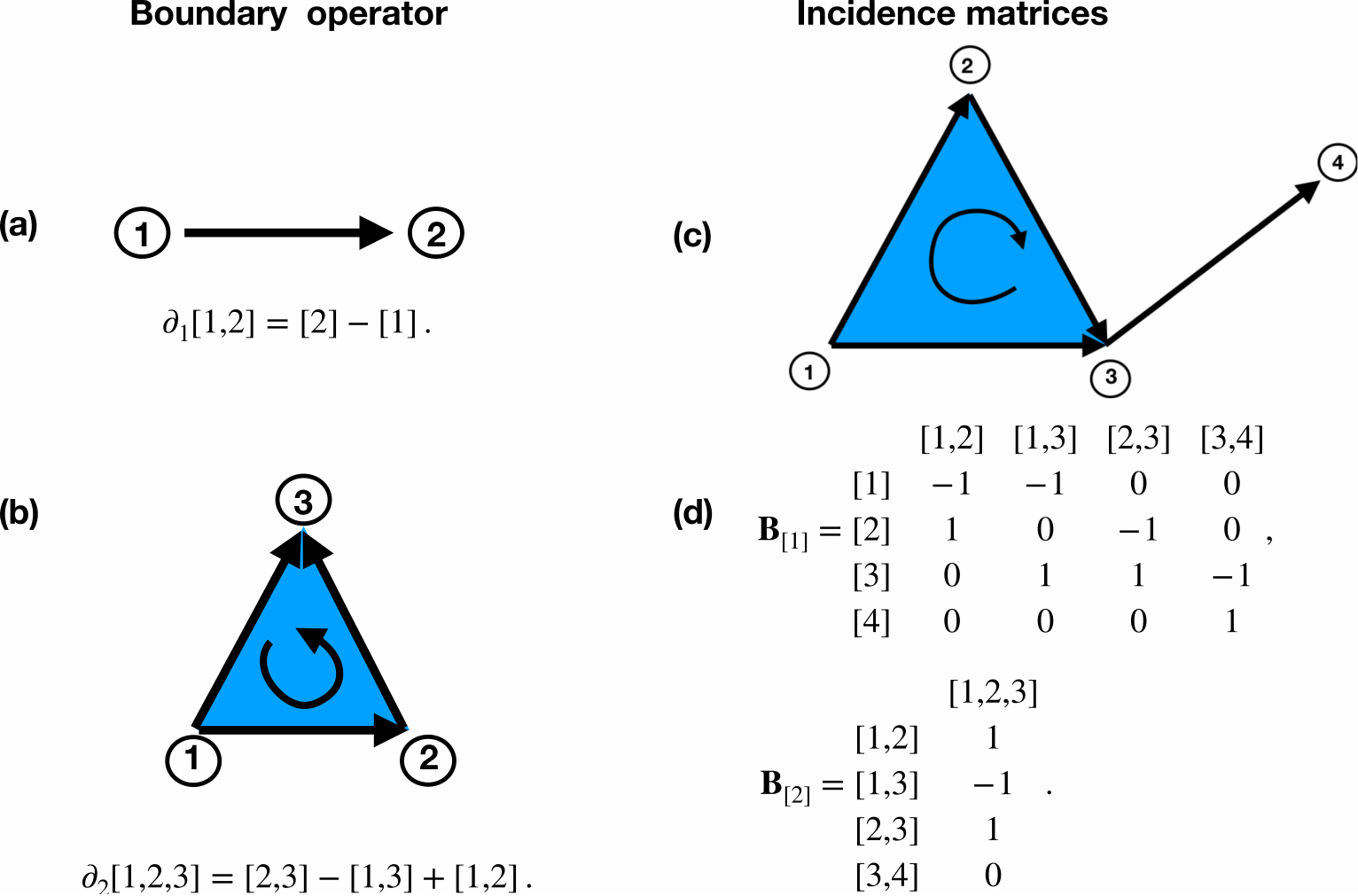}
\caption{{\bf The boundary operators and their representation in terms of the incidence matrices.} Panel (a) and (b) describe the action of the boundary operator on an oriented link and on an oriented triangle respectively. Panel (c) shows a toy example of a simplicial complex and panel (d) indicates its incidence matrices ${\bf B}_{[1]}$ and ${\bf B}_{[2]}$ representing the boundary operators $\partial_1$ and $\partial_2$ respectively.}
\label{fig:boundary}
\end{figure*}
\subsection{Oriented simplices and boundary map}

In algebraic topology simplices are oriented.
For instance a link $\alpha=[i,j]$ has the opposite sign of the link $[j,i]$, i.e.,
\bea
[i,j]=-[j,i].
\eea
Similarly to higher order simplices we associate an orientation such that 
\bea
[i_0,i_1,\ldots, i_n]=(-1)^{\sigma(\pi)}[i_{\pi(0)},i_{\pi(1)},\dots,i_{\pi(n)}],
\eea
where $\sigma(\pi)$ indicates the parity of the permutation ${\bf \pi}$.
It is good practice to use as orientation of the simplices the orientation induced by the labelling of the nodes, i.e., giving, for example, a positive orientation to any  simplex 
\bea
[i_0,i_1,\ldots, i_n],
\eea
where 
\bea
i_0<i_1<i_2\ldots<i_n.
\eea
This will ensure that the spectral properties of the higher-order Laplacians that will be defined later are independent of the labelling of the nodes.
 Given a simplicial complex, a  {\em $n$-chain} consists of the elements of a free abelian group $\mathcal{C}_n$ with basis  formed by the set of all oriented $n$-simplices. Therefore every element of $\mathcal{C}_n$ can be  uniquely expressed as a linear combination of the basis elements ($n$-simplices) with  coefficients in $\mathbb{Z}_2$.
The boundary operator $\partial_n$ is a linear operator $\partial_{n}:\mathcal{C}_n\to \mathcal{C}_{n-1}$ whose action is determined by the action on each $n$-simplex of the simplicial complex given by 
\bea
\partial_n [i_0,i_1\ldots,i_n]=\sum_{p=0}^n(-1)^p[i_0,i_1,\dots,i_{p-1},i_{p+1},\dots,i_n].
\label{boundary}
\eea
{As a concrete example,} in  Figure \ref{fig:boundary} we demonstrate the action of the boundary operator on links and triangles. 
A celebrated property of the boundary operator is that the {\em boundary of a boundary is null}, i.e.
\bea
\partial_{n}\partial_{n+1}=0
\label{bob}
\eea
for any $n>0$. This relation can be directly proven by using Eq.~(\ref{boundary}).
Let us consider a simplicial complex $\mathcal{K}$. Let us indicate with  $N_{[n]}$ the number of simplices of the simplicial complex with  generic dimension $n$.
Given a basis for the linear space of $n$-chains $\mathcal{C}_n$ and for the linear space of $(n-1)$-chains $\mathcal{C}_{n-1}$ formed by an ordered list of the $n$ simplices and $(n-1)$ simplices of the simplicial complex, the boundary operator $\partial_n$ can be represented  as  $N_{[n-1]}\times N_{[n]}$ incidence matrix ${\bf B}_{[n]}$. In Figure $\ref{fig:boundary}$ we show a $2$-dimensional simplicial complex and its corresponding incidence matrices ${\bf B}_{[1]}$ and ${\bf B}_{[2]}$.
Given that the boundary matrices obey Eq.~$(\ref{bob})$ it follows that the incidence matrices obey 
\bea
{\bf B}_{[n]}{\bf B}_{[n+1]}={\bf 0}, &{\bf B}_{[n+1]}^{\top}{\bf B}_{[n]}^{\top}={\bf 0},
\label{imo}
\eea
 for any $n>0$.
\subsection{Higher order Laplacians}
Using the incidence  matrices it is natural to generalize the definition of the graph Laplacian 
\bea
{\bf L}_{[0]}={\bf B}_{[1]}{\bf B}^{\top}_{[1]}
\eea
to  the higher-order Laplacian ${\bf L}_{[n]}$(also called combinatorial Laplacians) \cite{horak2013spectra,torres2020simplicial,barbarossa2020topological} that can be represented as a $N_{[n]}\times N_{[n]}$ matrix given by 
\bea
{\bf L}_{[n]}={\bf L}_{[n]}^{\text{down}}+{\bf L}_{[n]}^{\text{up}}
\eea
with 
\bea
{\bf L}_{[n]}^{\text{down}}&=&{\bf B}^{\top}_{[n]}{\bf B}_{[n]},\nonumber \\
{\bf L}_{[n]}^{\text{up}}&=&{\bf B}_{[n+1]}{\bf B}^{\top}_{[n+1]},
\eea
for $n>0$.  
The higher-order Laplacian can be proven to be independent of the orientation of the simplices as long as the simplicial complex has an orientation induced by a labelling of the nodes.

The most celebrated property of  higher-order Laplacian  is that the degeneracy of the zero eigenvalue of the $n$ Laplacian ${\bf L}_{[n]}$ is equal to the Betti number $\beta_n$ and that their corresponding eigenvectors localize around the corresponding $n$-dimensional cavities of the simplicial complex.
The higher-order Laplacians can be used to define higher-order diffusion \cite{torres2020simplicial} and can display {a higher-order} spectral dimension on network geometries.
Here we are particularly interested in the use of incidence matrices and higher-order Laplacians to define higher-order topological synchronization.

\subsection{Steady-state solution of the annealed equations for the NL Model}\label{steadystate}
Here we study Eqs.~(\ref{R0s_th}), (\ref{comp}) assuming that the distributions $g(\omega)$ and $G_i(\hat{\omega})$ are unimodal functions symmetric about their means.
Setting $\Psi = \hat \Psi = 0$ and considering the change of variables $z = {\omega }/({\sigma R_0 R_1^{\text{down}}})$, $y ={\hat \omega}/({\sigma R_0})$, Eqs.~(\ref{R0s_th}) can be written as
\begin{widetext}
\bea
1&=&\sigma  R_1^{\text{down}} \sum_{i=1}^N\frac{k_i^2}{\avg{k}N}\int_{-1}^1g(\Omega_0+z \sigma k_i \hat{R}_0R_1^{\text{down}})\sqrt{1-z^2}dz,\nonumber \\
R_0&=&\sigma \hat{R}_0 R_1^{\text{down}}\sum_{i=1}^N \frac{k_i}{N}\int_{-1}^1g(\Omega_0+z \sigma k_i \hat{R}_0R_1^{\text{down}})\sqrt{1-z^2}dz,\nonumber
\eea
while Eqs.~(\ref{comp}) reduce to 
\bea
R_1^{\text{down}}&=&\sigma R_0
\sum_{i=1}^N\frac{k_i}{N}\int_{-1}^1G_i(\avg{\hat{\omega}_i}+y \sigma R_0 k_i)\sqrt{1-y^2}d y,\nonumber\\
\hat{R}_1^{\text{down}}&=&\sigma R_0
\sum_{i=1}^N\frac{k_i^2}{\avg{k}N}\int_{-1}^1G_i(\Avg{\omega}_i+y \sigma R_0 k_i)\sqrt{1-y^2}d y.\nonumber
\eea
\end{widetext}
We notice that the equations for $R_0,\hat{R}_0$ and $R_1^{\text{down}}$ do not depend on the order parameter $\hat{R}_1^{\text{down}}$ so we can obtain a fully analytical solution of the model without solving the last equation. 
The above equations depend on the distribution $g(\omega)$ and the set of marginal distributions $G_i(\hat{\omega}_i)$.
However we can show that, provided $\avg{k^2}/\avg{k}$ is finite, the solution of these equations does not converge to the trivial solution $R_0=\hat{R}_0=R_1^{\text{down}} = 0$ for any finite value of $\sigma$.
Indeed  we are now going to show that the unstable branch of the solution these equations converges to the trivial solution only in the limit $\sigma\to\infty$.
Assuming $0<R_0\ll 1$, $0<\hat{R}_0\ll 1$ and $0<R_1^{\text{down}}\ll 1$ we can  expand the functions $g(z \sigma k_i \hat{R}_0R_1^{\text{down}})$ and $G_i(y \sigma R_0 k_i)$ as
\bea
g(\Omega_0+z \sigma k_i \hat{R}_0R_1^{\text{down}})&\simeq &g(\Omega_0)+\frac{g{\prime\prime}(\Omega_0)}{2}(z \sigma k_i \hat{R}_0R_1^{\text{down}})^2\nonumber \\
G_i(\Avg{\hat{\omega}_i}+y \sigma R_0 k_i)&\simeq &G_i(\Avg{\hat{\omega}_i})+\frac{G_i^{\prime\prime}(\Avg{\hat{\omega}_i})}{2}(y \sigma R_0 k_i)^2
\nonumber\eea
Stopping at the first order of this expansion we get
\bea
1=\sigma R_1^{\text{down}}g(\Omega_0)\frac{\pi}{2}\frac{\Avg{k^2}}{\Avg{k}}, \\
R_0=\sigma \hat{R_0}R_1^{\text{down}}g(\Omega_0)\frac{\pi}{2}\Avg{k},\\
R_1^{\text{down}}=\sigma\hat{R}_0\frac{\pi}{2}\frac{1}{N}\sum_i k_i G_i(\Avg{\hat{\omega}_i}).
\eea
This equations lead to the following scaling of $R_0$ and $R_1^{\text{down}}$ with $\sigma$
\bea
R_0&=&\sigma^{-2}J_0,\nonumber \\
R_1^{\text{down}}&=&{\sigma}^{-1}J_1,
\label{AsyAp}
\eea
with 
\bea
J_0&=&\Avg{k}\left[\frac{\pi}{2}\frac{\Avg{k^2}}{\Avg{k}}\right]^{-2}\left[g(\Omega_0)\frac{1}{N}\sum_i k_i G_i(\Avg{\hat{\omega}_i})\right]^{-1},\nonumber \\
J_1&=&\left[g(\Omega_0)\frac{\pi}{2}\frac{\Avg{k^2}}{\Avg{k}}\right]^{-1}.
\label{Jap}
\eea
This confirms the theoretical framework revealing that in this dynamics there is always a trivial solution $R_0=\hat{R}_0=R_1^{\text{down}}=0$ while Eqs.~(\ref{R0s_th}), (\ref{comp}) are characterized by a saddle-point instability so that   for $\sigma>\sigma_c$ two additional solutions emerge, a stable solution and an unstable solution. The stable solution describes the synchronized phase and captures the backward transition. As long as the second moment of the degree distribution does not diverge, the unstable solution  converges to the trivial solution $R_0=\hat{R}_0=R_1^{\text{down}}=0$ only for $\sigma\to\infty$.

The asymptotic scaling for $R_0$ and $R_1^{\text{down}}$ given by Eq.~(\ref{AsyAp}) can be adapted to capture the asymptotic scaling of the fully connected case with a suitable rescaling of the model parameters of the model, obtaining  Eqs.~(\ref{AsyFC}), (\ref{JFC}).

\subsection{Marginal distributions in the fully connected case}\label{Ap3}
The distribution $G_1(\hat{\bm\omega})$ of  $\hat{\bm\omega}$ is a Gaussian distribution with averages given by Eq. (\ref{avgomega}) and covariance matrix $\bf C$ given by Eq.(\ref{correlation}). 
The covariance matrix has $N-1$ eigenvalues given by $\lambda=1/\tau_1^2$ and one zero eigenvalue $\lambda=0$ corresponding to the eigenvector
\bea
{\bf 1}/\sqrt{N}=(1,1,\ldots, 1)^{\top}/\sqrt{N}.
\eea
This means that we should always have 
\bea
\sum_{n=1}^N\frac{[\hat{\omega}_n-\Avg{\hat{\omega}_n}]}{\sqrt{N}}=0,
\eea
a constraint that we can introduce as a delta function in the expression for the joint distribution $\hat{G}(\hat{\bm\omega})$ of the vector $\hat{\bm\omega}$.
Here we note that under these hypotheses and assuming that the distribution of the frequencies of the links is a Gaussian with average $\Omega_1/N$ and standard deviation $1/(\tau_1\sqrt{N-1})$  the marginal probability  $G_i(\hat{\omega})$ of $\hat{\omega}_i$ can be expressed as Eq.~(\ref{marginal}).

Given that the covariance matrix has a zero eigenvalue we can express the joint Gaussian distribution $\hat{G}(\hat{\bm\omega})$ as 
\bea
\hat{G}(\hat{\bm\omega})={\mathcal{C}}e^{-\mathcal{F}(\hat{\bm\omega})}\delta\left(\sum_{n=1}^N\frac{[\hat{\omega}_n-\Avg{\hat{\omega}_n}]}{\sqrt{N}}\right),
\label{G1a3}
\eea
where $\delta(x)$ indicates the delta function and where $\mathcal{F}(\hat{\bm\omega})$  and $\mathcal{C}$ are given by 
\bea
\mathcal{F}(\hat{\bm\omega})&=&\frac{\tau_1^2}{2}\sum_{n=1}^{N}\left(\hat{\omega}_n-\Avg{\hat{\omega}_n}\right)^2,\nonumber \\
\mathcal{C}&=&\left(\frac{\tau_1}{\sqrt{2\pi}}\right)^{N-1}.
\eea
The marginal probability $G_i(\hat{\omega})$ is given by 
\bea
G_i(\hat{\omega})=\int \prod_{n\neq i}d\hat{\omega}_n \hat{G}(\hat{\bm\omega}).
\eea
By expressing the delta function in Eq. (\ref{G1a3}) in its integral form
\bea
\delta({x,y})=\frac{1}{2\pi}\int_{-\infty}^{\infty} dz e^{\mathbb{i}z (x-y)}
\eea
we get  the final expression for the marginal distribution Eq.~(\ref{marginal}), in fact we have

\begin{widetext}
\bea
G_1^{(i)}(\hat{\omega})&=&\frac{\mathcal{C}}{2\pi}\int dz\int \prod_{n\neq i}d\hat{\omega}_n e^{-\mathcal{F}(\hat{\bm\omega})}\exp\left[\mathbb{i}z\left(\sum_{n=1}^N\frac{[\hat{\omega}_n-\Avg{\hat{\omega}_n}]}{\sqrt{N}}\right)\right]\nonumber \\
&=&\frac{e^{-\tau_1^2\frac{[\hat{\omega}_i-\Avg{\hat{\omega}_i}]}{2}}}{2\pi}\int dz \exp\left[-\frac{z^2}{2\tau_1^2\bar{c}}+\mathbb{i}z\frac{[\hat{\omega}_i-\Avg{\hat{\omega}_i}]}{\sqrt{N}}\right]\nonumber \\
&=& \frac{\tau_1}{\sqrt{2\pi/\bar{c}}}\exp\left[-\tau_1^2\bar{c}\frac{(\hat{\omega}_i-\Avg{\hat{\omega}_i})^2}{2}\right].
\eea
\end{widetext}

\end{document}